\documentclass[pra,aps,twocolumn,reprint,floatfix,noeprint]{revtex4-1}
\usepackage[draft,authormarkup=none]{changes}
\setaddedmarkup{\color{red}{#1}}
\usepackage[utf8]{inputenc}
\usepackage{graphicx,amssymb,amsmath}
\usepackage{amsmath}
\usepackage{amsfonts}
\usepackage{braket}
\usepackage{txfonts}
\usepackage{hyperref}
\usepackage{bbm}
\usepackage{mathbbol,times}
\def\Tr{\hbox{Tr}}

\renewcommand{\vec}{\mathbf}
\let\originalleft\left
\let\originalright\right
\renewcommand{\left}{\mathopen{}\mathclose\bgroup\originalleft}
\renewcommand{\right}{\aftergroup\egroup\originalright}
\renewcommand{\right}{\aftergroup\egroup\originalright}
\begin{document}
\title{Generation of Coherence via Gaussian Measurements}
\author{Francesco Albarelli}
\email{francesco.albarelli@unimi.it}
\homepage{http://users.unimi.it/aqm}
\affiliation{Quantum Technology Lab,
Dipartimento di Fisica, Universit\`a degli Studi di Milano,
20133 Milano, Italy}
\author{Marco G. Genoni}
\email{marco.genoni@fisica.unimi.it}
\homepage{http://users.unimi.it/aqm}
\affiliation{Quantum Technology Lab,
Dipartimento di Fisica, Universit\`a degli Studi di Milano,
20133 Milano, Italy}
\author{Matteo G. A. Paris}
\email{matteo.paris@fisica.unimi.it}
\homepage{http://users.unimi.it/aqm}
\affiliation{Quantum Technology Lab, Dipartimento di Fisica,
Universit\`a degli Studi di Milano, 20133 Milano, Italy}
\affiliation{INFN, Sezione di Milano, I-20133 Milano, Italy}
\date{\today}
\begin{abstract}
We address measurement-based generation of quantum coherence in 
continuous variable systems. We consider Gaussian measurements 
performed on Gaussian states and focus on two scenarios: 
In the first one, we assume an initially correlated 
bipartite state shared by two parties, and study how correlations 
may be exploited to remotely create quantum coherence via measurement 
back-action. In particular, we focus on conditional states with zero 
first moments, so as to address coherence due to properties of the 
covariance matrix. We consider different classes of bipartite states 
with incoherent marginals and show that the larger the {\em 
measurement squeezing}, the larger the conditional coherence. 
Homodyne detection is thus the optimal Gaussian measurement to 
remotely generate coherence. 
We also show that for squeezed thermal states there exists a 
threshold value for the generated coherence which separates 
entangled and separable states at a fixed energy. 
Finally, we briefly discuss the tripartite case and the 
relationship between tripartite correlations and the conditional
two-mode coherence. In the second 
scenario, we address the steady state coherence of a system interacting 
with an environment which is continuously monitored. In particular, 
we discuss the dynamics of an optical parametric oscillator
in order to investigate how the coherence 
of a Gaussian state may be increased by means of time-continuous
Gaussian measurement on the interacting environment. 
\end{abstract}
\maketitle
\section{Introduction}
Coherence is the main ingredient to observe interference, a physical 
phenomenon at the basis of several applications in science and technology. 
In addition, since the superposition principle is the main point of 
departure from classical physics, coherence is also at the basis of 
purely quantum features, such as entanglement, and it plays a pivotal
role in the description of quantum states and operations. In turn, coherence 
is a key concept in several fields, ranging from quantum optics\cite{Glauber1963a,MANDEL1965,Albrecht1994} to quantum information\cite{Hillery2015,Shi2016}, quantum thermodynamics\cite{Lostaglio2015,Korzekwa} and quantum estimation\cite{Giorda2016}, as well as in several phenomena in biological systems\cite{Huelga2013a}.
\par
Even if quantum coherence has long been recognized as a resource in many 
contexts, a systematic  study from a resource-theoretical point 
of view has started only recently\cite{Baumgratz2014} and it is undergoing 
very active development (see~\cite{Streltsov2016a} for a recent review).
In turn, the generation and manipulation of coherence in bipartite and 
multipartite systems, as well as the 
interplay with quantum correlations, are topics which have recently
gained considerable attention\cite{Ma2016,Ma2016a,Chitambar2016,Wu2017,Hu2016a,Hu2016a,Li2016c,Zhang,Wang2017b}.
Our aim here is to explore these connections in continuous variable systems and 
in particular for Gaussian states and operations. As a matter of fact, despite quantum coherence having been discussed mostly for finite-dimensional systems, a resource-theoretic framework for coherence for infinite dimensional Fock space has been introduced\cite{Zhang2015d} and the coherence of Gaussian states has been studied in some detail\cite{Xu2015c,Buono2016}. We also mention that for 
continuous variable systems, the coherence in the non-orthogonal coherent 
states basis has recently been investigated\cite{Gehrke2012,Ryl2017,Tan2017} and linked 
to $P$ function nonclassicality.
\par
In this paper, we address the issue of generating  quantum coherence in the 
Fock basis by performing Gaussian measurements on Gaussian states. We 
analyze two scenarios in detail. In the first case, we assume an initially correlated bipartite state shared by two distant labs. 
The two marginal states are initially incoherent and we analyze how 
correlations can be exploited to remotely create quantum coherence via 
measurement back-action. In the second case, we study the coherence of
the steady state of a system interacting with an environment which is 
continuously monitored, i.e. subject to continuous-time Gaussian 
measurements. Also in this case, correlations between the environment and 
the system, which are provided by the dynamics, play a role in the generation
of coherence. In order to address only the coherence due to the covariance 
matrix (CM), i.e., to correlations, most of the results are obtained neglecting the 
first moments. However, for the sake of completeness, some results including 
first moments are reported in the Appendix.
\par
The paper is structured as follows. In Sec.~\ref{sec:Gaussian} we briefly 
recall the formalism for bosonic Gaussian states and operations and set the 
notation. In Sec.~\ref{sec:Coh_Meas}, we review coherence measures which 
are suitable for Gaussian states also highlighting some connections between 
quantum correlations and quantum coherence. Section~\ref{sec:remote_creation} 
is devoted to the (conditional) remote creation of quantum coherence 
starting either from generic two-mode Gaussian states in normal form or from a 
specific class of three-mode states. Sec.~\ref{sec:continuous} 
deals with coherence enhancement due to continuous measurements on the 
environmental modes. In particular, we focus on the example 
of a monitored parametric oscillator. Finally, Sec.~\ref{sec:conclusions} closes 
the paper with some concluding remarks.
\section{Gaussian states and measurements}
\label{sec:Gaussian}
We consider a set of $n$ bosonic modes described by a vector of quadrature 
operators $\hat{\bf r}^{\sf T}=(\hat{x}_1,\hat{p}_1,\dots,\hat{x}_n,\hat{p}_n)$, that satisfy the canonical commutation relation $[\hat{r}_j, \hat{r}_k] = i \Omega_{jk}$, where $\Omega$ is the symplectic matrix 
\begin{equation}
\Omega = \bigoplus_{j=1}^{n} \omega, \quad \text{with} \quad \omega = \begin{pmatrix} 0 & 1\\ -1 & 0 \end{pmatrix}.
\end{equation}
A quantum state $\varrho$ is defined as Gaussian if and only if can be written 
as a ground or thermal state of a quadratic Hamiltonian, {\em i.e.} 
\begin{align}
\varrho = \frac{\exp\{-\beta \hat{\mathcal{H}}_{\sf G}\}}{Z} \,, \:\:\: \beta \in \mathbbm{R}\:,
\end{align}
where $\hat{\mathcal{H}}_{\sf G}= \hat{\bf r}^{\sf T} H_G \,\hat{\bf r}/2$ 
and $H_G \geq 0$ \cite{Genoni2016}. Gaussian states can be univocally 
described by the vector of first moments $\bar{\bf r}$ and the covariance 
matrix $\sigma$:
\begin{align}
\bar{\bf r} = \Tr[ \varrho \hat{\bf r}]\:,  \qquad
\sigma_{ij} = \Tr[\varrho \{ (\hat{r}_i - \bar{r}_i) , (\hat{r}_j - \bar{r}_j)^{\sf T} \} ] \:.
\end{align}
In order to describe a proper Gaussian quantum state, the CM has to satisfy the physicality condition \cite{Simon1994}:
\begin{align}
\sigma + i \Omega \geq 0. \label{eq:physical}
\end{align}
A generic bipartite Gaussian state is completely described by a 
first moment vector and a block-form CM
\begin{equation}
\label{eq:STS}
\bar{\bf r} = \begin{pmatrix} \bar{\bf r}_A \\ \bar{\bf r}_B \end{pmatrix} \qquad 
\sigma=\begin{pmatrix}
\sigma_A  & \sigma_{AB} \\
\sigma_{AB}  & \sigma_B
\end{pmatrix},
\end{equation}
where $\bar{\bf r}_A$, $\sigma_A$ and $\bar{\bf r}_B$, $\sigma_B$ are the first moments and CMs of the marginal states, while $\sigma_{AB}$ contains the correlations betweens the two subsystems. A necessary and sufficient condition for the entanglement of a bipartite Gaussian state has been derived in~\cite{Simon2000}, by applying the physicality condition~\eqref{eq:physical} to the covariance matrix of the partially transposed state.
\par
In this paper, we deal with Gaussian measurements on bipartite 
Gaussian states. A generic Gaussian measurement is described by 
a probability operator-valued measure (POVM) of the form 
$\Pi(\mathbf{r}) = D(\mathbf{r}) \varrho_G D^\dag (\mathbf{r})$ where 
$\varrho_G$ is a Gaussian state and $D(\mathbf{r})$ is the 
displacement operator. Any Gaussian measurement
is thus naturally associated with 
the CM $\sigma_m$ of a Gaussian state, and, in 
particular, measurements involving ideal detectors with no losses 
correspond to pure 
states such that $\det \sigma_m = 1$. Gaussian measurements 
include the case of homodyne and heterodyne detections, whose 
CMs for single-mode detection read
\begin{align}
\sigma_m^{\sf (hom)}= \lim_{s \rightarrow 0}\,
R(\phi)\, \begin{pmatrix}
s  & 0 \\
0  & s^{-1} 
\end{pmatrix}\, R^{\sf T} (\phi)\,, \qquad
\sigma_m^{\sf (het)}= \mathbbm{1}_2 \,,
\end{align}
respectively, where $R(\phi)$ denotes a real rotation matrix.
\par
Given an initial bipartite Gaussian state, with moments given in (\ref{eq:STS}), 
the conditional state for the subsystem $A$ after a Gaussian measurement 
$\sigma_m$ on subsystem $B$, has the following covariance matrix 
and first-moments vector~\cite{Genoni2016,bla12}
\begin{equation}
\begin{split}
\label{eq:cond_state}
\sigma'_{A} & =  \sigma_A  -\sigma_{AB} \left(\sigma_B + \sigma_m \right)^{-1} \sigma_{AB}^{\intercal} \\
\vec{r}'_{A} & = \vec{r}_A + \sigma_{AB} \left(\sigma_B + \sigma_m \right)^{-1} (\vec{r}_{\text{out}} - \vec{r}_B),
\end{split}
\end{equation}
where the vector $\vec{r}_{\text{out}}$ is the vector of the outcomes, which 
are distributed according to a Gaussian centered at $\vec{r}_B$:
\begin{equation}
\label{eq:p_rout}
p(\vec{r}_{\text{out}}) = \frac{e^{-(\vec{r}_{\text{out}} - \vec{r}_B)^{\intercal} (\sigma_m + \sigma_B)^{-1} (\vec{r}_{\text{out}} - \vec{r}_B)}}{\pi \sqrt{\det(\sigma_m + \sigma_B)}}.
\end{equation}
\section{Coherence Measures for Gaussian States}
\label{sec:Coh_Meas}
\subsection{Resource theory of coherence in the Fock space}
We consider the infinite-dimensional Hilbert space of a single-mode bosonic system and 
the Fock basis $\left\{ | n \rangle \right\}_{n=0}^\infty$ as the reference basis to 
assess the coherence of a state. 
The Fock states are defined as the eigenstates $H|n\rangle 
= n |n\rangle$ of the harmonic oscillator free Hamiltonian
$H = \frac12\left(\hat{x}^2 + \hat{p}^2 - 1\right)$.
This is the most natural choice for discussing quantum coherence of bosonic continuous variable states~\cite{Zhang2015d,Xu2015c,Buono2016}.
\par 
We denote by $\mathcal{I}$ the set of incoherent states $\delta = 
\sum_n \delta_n |n \rangle \langle n |$ with $\sum_n \delta_n = 1$, 
where all the sums run from 0 to $\infty$. 
Incoherent quantum operations $\phi_{\text{ICPTP}}$~\cite{Baumgratz2014} are 
completely-positive and trace-preserving (ICPTP) maps $\phi_{\text{ICPTP}}
\left( \circ \right) = \sum_n K_n \circ K_n^{\dag}$, $\sum_n K_n K_n^\dag = 
\mathbb{1}$ for which $K_n \mathcal{I} K_n^\dag \subset \mathcal{I}$, i.e. 
the Kraus operators of ICPTP maps send incoherent states to incoherent states. 
This is the resource theoretical framework we use throughout the 
paper. Notice that different definitions of incoherent operations may be 
employed, which leads to different resource theories~\cite{Chitambar2016b,Chitambar2016c,Marvian2016}.
\par
Any coherence measure functional $C$ should satisfy the following properties.
\begin{itemize}
\item[(C1)] $C(\rho) \geq 0 \quad \forall \rho$ with $C(\rho)=0 \iff \rho \in \mathcal{I}$;
\item[(C2a)] monotonicity under ICPTP operations, $C(\rho) \geq C\left( \phi_{\text{ICPTP}}  \left( \rho \right) \right)$;
\item[(C2b)] monotonicity under selective measurements, on average, $C(\rho) \geq \sum_n p_n C(\rho_n) \quad \forall K_n$, with $\rho_n = K_n \rho K_n^\dag / p_n$ and $p_n = \Tr \left[ K_n \rho K_n^\dag \right]$;
\item[(C3)] convexity, i.e., $C$ is nonincreasing under mixing, $\sum p_n C(\rho_n) \geq C\left
(\sum_n p_n \rho_n \right)$.
\end{itemize}
Furthermore, for states of an infinite dimensional system, we require that 
states with finite energy (i.e., a finite average number of bosonic excitations) 
have a finite coherence~\cite{Zhang2015d}
\begin{itemize}
\item[(C4)] $\Tr \left[ \rho \, \hat{n} \right] < \infty \implies C ( \rho ) < \infty$.
\end{itemize}
We point out that requirements (C2b) and (C3) are equivalent to the additivity of coherence 
for block diagonal density operators in the reference basis~\cite{Yu2016}.
\par
A class of coherence measures is then obtained by minimizing any pseudo-distance of the quantum state under investigation from the set of incoherent states $\mathcal{I}$. A convenient choice is given by the quantum relative entropy 
$S(\rho || \sigma ) = -\Tr [ \rho \log \sigma] - S(\rho) $, where 
$S(\rho)=-\Tr \left[ \rho \log \rho \right]$ is the Von Neumann entropy 
of the density matrix $\rho$. The resulting measure is the so-called 
relative entropy of coherence 
\begin{align}
C_{S} = \min_{\delta \in \mathcal{I}} S( \rho || \delta ) = S( \rho || \rho_{\text{diag}} ) = S(\rho_{\text{diag}}) - S(\rho) \,,
\end{align}
where $\rho_{\text{diag}}$ is the original state with all the off-diagonal elements in the reference basis suppressed. This measure satisfies all properties (C1)-(C3) and crucially also property (C4)~\cite{Zhang2015d}. As a consequence, it 
is a good coherence monotone for infinite dimensional systems.
\par
The explicit expression of the relative entropy of coherence in 
the Fock basis is given by
\begin{equation}
\label{eq:coh_rel_ent1}
C_S \left( \rho \right) = H \left( \left\{ p_n \right\} \right) - S(\rho),
\end{equation}
where $\left\{ p_n = \langle n | \rho | n \rangle \right\}$ is the photon number distribution 
and $H( \left\{ p_n \right\} )= - \sum_{i=0}^{\infty} p_n \log (  p_n  )$ is the classical 
Shannon entropy. This expression has the drawback that $H( \left\{ p_n \right\} )$ is not easy 
to obtain in closed form even in the case of Gaussian states. We also notice 
that the relative entropy of coherence may also be expressed in terms of the 
entropic measure of non-Gaussianity $\delta(\rho)$ \cite{ng1,ng2} as follows
\begin{align}
C_S \left( \rho \right) = \delta(\rho) + H \left( \left\{ p_n \right\} \right) - 
h \left( \sqrt{\det\sigma } \right)\,, \label{csng}
\end{align}
where $\sigma$ is the covariance matrix of $\rho$ and $h(x)=\left(\frac{x+1}{2}\right)\log \left(\frac{x+1}{2}\right)-\left(\frac{x-1}{2}\right)\log \left(\frac{x-1}{2}\right)$
\par
The relative entropy of coherence can be straightforwardly extended to 
multimode Fock space~\cite{Zhang2015d}; for example for a two-mode state it reads
\begin{equation}
\label{eq:coh_rel_ent_2mode}
C_S(\rho_{AB})= H ( \left\{  p_{nm} \right\} ) - S(\rho_{AB}),
\end{equation}
with $p_{nm} = \langle n, m | \rho_{AB} | n,m \rangle 	$.
\subsection{Gaussian coherence}
The resource theory of coherence has also been studied focusing 
on Gaussian states. In this case one is interested in ICPTP operations 
which preserve the Gaussian character of the state~\cite{Xu2015c}. For 
single mode systems the set of incoherent Gaussian states 
$\mathcal{I_G}$ only includes thermal states~\cite{Xu2015c,Buono2016} 
(which we label with the Greek letter $\nu$). For multimode systems, 
the only incoherent states are locally thermal states (tensor products
 of thermal states), i. e. $\otimes_{i=1}^{m} \nu_i$ for an $m$-mode state~\cite{Xu2015c}, whose covariance matrix is a direct sum of multiples of identity matrices $\sigma_\nu = \oplus_{i=1}^m k_i \mathbbm{1}_2$.\\
The Gaussian relative entropy of coherence is thus obtained by restricting 
the minimization to the set of incoherent Gaussian states $\mathcal{I_G}$~\cite{Xu2015c}:
\begin{equation}
\label{eq:coh_rel_ent2}
C_{S}^{G} (\rho_G) = 	\inf_{\nu} \left\{ S(\rho_G || \nu) \, |  \, \nu \in \mathcal{I_G} \right\}\,.
\end{equation}
This is  an upper bound to the relative entropy of coherence, as the closest incoherent state need not be Gaussian. 
For a single mode the closest Gaussian incoherent state 
is a thermal state with the same mean photon number, leading to
\begin{align}
C_{S}^{G} (\rho_G) &= \min_{\nu \in \mathcal{I}_\mathcal{G}} S(\rho_G || \nu ) \nonumber \\
&= S(\rho_G || \bar{\nu}) = S(\bar{\nu}) - S(\rho_G) \\ 
& =  h\left(2 \bar{n} +1\right) - h\left(\sqrt{\det \sigma}\right) \label{csg} \\
& = h \left( \frac{1}{2} \Tr \left[ \sigma \right] + | \vec{r} |^2 \right) - h \left( \sqrt{\det \sigma}\right) \label{eq:coh_rel_ent2b}  ,
\end{align}
where $\bar{\nu}$ is the thermal state with $\bar{n} = \Tr [\hat{n} \, \rho_G ] = \frac{1}{4} \Tr \left[ \sigma \right] + \frac{1}{2} | \vec{r} |^2 - \frac{1}{2}$ thermal photons, and $\sigma$ and $\vec{r}$ are the covariance matrix and first moments vector of the Gaussian state $\rho_G$. Expression (\ref{csg}) follows
from Eq. (\ref{csng}) upon noticing that for Gaussian states $\delta(\rho)
\rightarrow 0$ and $H \left( \left\{ p_n \right\} \right) \rightarrow H 
\left( \left\{ \nu_n \right\} \right) \equiv h\left(2 \bar{n} +1\right)$.
Also the Gaussian coherence may be generalized to $m$ modes as
\begin{equation}
	C_{S}^{G} (\rho_G) = \sum_{i=1}^m S(\bar{\nu}_i) - S(\rho_G),
\end{equation}
where the $\bar{\nu}_i$ are single-mode thermal states at the energy of the $i$th mode, i.e. $\bar{n}_i = \frac{1}{4} \left( \Tr \left[ \sigma_i \right] + 2 | \vec{r}_i |^2 - 2 \right)$.
Coherence measures based on proper geometrical distances, such as Bures 
and Hellinger distances, have also been investigated~\cite{Buono2016}. 
In the present work we focus on measures based on the relative entropy.
\subsection{Coherence and correlations}
There are tight relationships between quantum coherence and correlations~\cite{Xi2014,Yao2015,Ma2016,Tan2016,Li2016c,Guo2016b}. Here we review and 
highlight some of these connections for bipartite Gaussian states.
In a bipartite system with two local reference bases $\left\{ |n\rangle_A \right\}$ and $\left\{ |n\rangle_B \right\}$, the key quantity 
is the difference $\Delta C$ between the total coherence in the tensor product basis $\left\{ |n\rangle_A \otimes | m \rangle_B \right\}$ and the local coherences. This quantity is also known as the correlated coherence~\cite{Tan2016,Guo2016b}. Using the entropic measure of coherence $C_S$ we have:
\begin{align}
\label{eq:DeltaCS}
\Delta C_S \left( \rho_{AB} \right) =  \: &  C_S \left( \rho_{AB} \right) - 
\left[ C_S \left(\rho_{A}\right) + C_S \left( \rho_{B} \right) \right]  \\
= \: & H ( \left\{  p_{nm} \right\} ) - S(\rho_{AB}) - H ( \left\{  p_{n} \right\} ) - H ( \left\{  p_{m} \right\} )\notag \\ & + S(\rho_A) + S(\rho_B)\\
=\:  & \mathcal{I}_q \left( \rho_{AB} \right) - \mathcal{I}\left( A : B \right);
\end{align}
i.e. $\Delta C_S$ is equal to the difference between the quantum mutual 
information $\mathcal{I}_q \left( \rho_{AB} \right)=S(\rho_A) + S(\rho_B)-S(\rho_{AB})$ and the classical mutual information 
$\mathcal{I}\left( A : B \right) = H ( \left\{  p_{n} \right\} ) 
+ H ( \left\{  p_{m} \right\}) - H ( \left\{  p_{nm} \right\} )$
of a channel based on measurements in the reference basis.
\par
The correlated coherence $\Delta C$ (independently of the coherence measure) has 
been introduced as basis-independent measure of quantum correlations~\cite{Tan2016} by fixing the eigenbases of the marginal states as a reference so that $C_S(\rho_A)=C_S(\rho_B)=0$. In particular, when the relative entropy of coherence 
is used, this measure corresponds to the measurement-induced disturbance (MID)~\cite{Luo2008}, denoted as $\mathcal{M}(\rho_{AB})$. From \eqref{eq:DeltaCS} we 
then have that MID is an upper bound to the correlated coherence: $ \Delta C_S ( \rho_{AB} ) \leq \mathcal{M} ( \rho_{AB}) $.
Another measure of quantum correlations, the ameliorated measurement-induced disturbance $\mathcal{A}(\rho_{AB})$ (AMID), is obtained by minimizing the 
classical mutual information over all possible local POVMs. 
Crucially AMID is an upper bound to the (entropic) quantum discord 
$\mathcal{M}(\rho_{AB}) \geq \mathcal{A}(\rho_{AB}) 
\geq \max \left[ D(A:B),D(B:A) \right]=M_D$, where 
$D(A:B)$ is the asymmetrical discord obtained by measuring subsystem $B$~\cite{Mista2011}.
Given the minimization over all possible measurements in the definition of $\mathcal{A}$ we have the following inequalities
\begin{align}
\label{eq:DeltaCineq}
M_D  \leq \mathcal{A} \left( \rho_{AB} \right) &\leq \Delta C_S \left( \rho_{AB} \right) \leq C_S \left( \rho_{AB} \right) \,.
\end{align}
We thus conclude that the 
relative entropy of coherence $C_S$ on the tensor product of local bases 
of a bipartite system is an upper bound to the discord. This is in 
complete analogy with discrete variable systems~\cite{Yao2015}, where the 
same result have been obtained by resorting to a geometric measure of 
quantum discord.
\par
Furthermore, if we consider the Gaussian relative entropy of coherence, then the quantity $\Delta C_S^G$ is equal to the quantum mutual information~\cite{Zheng2016}:
\begin{align}
\Delta C_{S}^{G} \left( \rho_{AB} \right) = \: & C_{S}^{G} (\rho_{AB}) - \left( C_{S}^{G} (\rho_{A})  + C_{S}^{G} (\rho_{B}) \right) \\
= \: & S(\bar{\nu}_{A}\otimes \bar{\nu}_B) - S(\rho_{AB}) \notag \\ 
& - \left[ S(\bar{\nu}_A) - S(\rho_A) + S(\bar{\nu}_B) - S(\rho_B)  \right]  \\
= \: & S(\rho_A) + S(\rho_B) - S(\rho_{AB}) = \mathcal{I}_q(\rho_{AB})\,.
\end{align}
Overall, we obtain a further (loose) bound to the quantum discord in terms of Gaussian quantities, expressed by the chain of inequalities 
\begin{align}
M_D  \leq \Delta C_S \left( \rho_{AB} \right) \leq \Delta C_S^G \left( \rho_{AB} \right) \leq C_S^G \left( \rho_{AB} \right)\,. \label{eq:bounddiscord}
\end{align}
\section{Remote Creation of Coherence}
\label{sec:remote_creation}
We now focus on the problem of remote creation of quantum coherence. 
In this scheme, we assume to have a correlated bipartite Gaussian 
state $\rho_{AB}$ and we want to study the quantum coherence generated 
on subsystem $A$ by performing Gaussian measurements on subsystem $B$. 
The term \emph{remote} comes from the fact that the marginal states 
$\rho_A$ and $\rho_B$, initially incoherent, may be manipulated at 
distant labs and one generates coherence on, say, system $B$ by performing
measurements on system $A$.
\par
We first investigate the intuitive idea that performing {\em squeezed} 
measurements may induce coherence on an initially incoherent marginal state. 
The possibility of creating coherence is due to the subsystems being correlated 
(not necessarily entangled). Therefore, we also study the interplay between the correlations and the remotely obtainable coherence. We show that remotely induced coherence can be used for entanglement detection, given the local energies or purities. A similar result has recently been obtained for extractable work with Gaussian measurements~\cite{Brunelli2017}.
\par
We mainly focus on two-mode states, but we also report an example of 
a feasible three mode state, to explicitly show that Gaussian measurements 
on one mode induce both coherence and correlations in the remaining modes. 
Similar features have been investigated in finite dimensional systems~\cite{Hu2016a,Zhang} and also generalized to arbitrary quantum operations 
beyond measurements~\cite{Ma2016a}.
\subsection{General considerations for two-mode systems}
At variance with the study of quantum correlations, the study of quantum 
coherence is highly influenced by local unitary operations. As a matter of 
fact, local displacement operations may increase the coherence of Gaussian 
states~\cite{Buono2016,Xu2015c} and, in turn, the first moments $\vec{r}$ 
may play a role. On the other hand, in order to point out the role (and the interplay) of correlations and measurement back-action in the
generation of coherence, we assume 
vanishing first moments in the initial bipartite state. For the same reasons, 
we focus on the coherence of the most-likely conditional state i.e.,
according to the probability of outcomes in Eq. (\ref{eq:p_rout}), the 
state with zero first moments. Indeed, the coherence gained by exploiting the first moments cannot be linked to quantum correlations, since the first moments of a bipartite 
state can be controlled by local operations only.  However, for completeness, 
in Appendix~\ref{sec:appendix} we also extend the analysis by taking into account 
the effect of first moments.
\par
We assume bipartite Gaussian states in normal form: states with zero mean 
$\vec{r}_A = \vec{r}_B = \left( 0, 0 \right)$ and with the 
 submatrices in Eq.~\eqref{eq:STS} all diagonal and parametrized as: 
 $\sigma_{A} = a \mathbb{1}_2$, $\sigma_{B} = b \mathbb{1}_2$ and $\sigma_{AB}=\mathrm{diag}(c_1,c_2)$. This choice is justified for two main reasons: (i) 
 we want to focus on a class of bipartite states with incoherent (thermal) 
 marginal states and (ii) as previously explained for local displacement 
 operations, also a local squeezing operation can indeed affect the coherence properties of the state, but it does not play any role as regards the 
 correlations that are the main focus of this study. 
\par
Without loss of generality, we choose a Gaussian measurement represented 
by a diagonal covariance matrix $\sigma_m = \mathrm{diag}\left(s,1/s\right)$, 
i.e. squeezed along the $x$ or $p$ direction. Making these assumptions 
the conditional state has the CM and first moments
\begin{equation}
\label{eq:cond_state2modes}
\begin{split}
\sigma'_{A} & = 
\begin{pmatrix}
a - \frac{c_1^2}{b+s} & 0 \\ 
0 & a - \frac{c_2^2}{b+s^{-1}}
\end{pmatrix} \\
\vec{r}'_{A} & = \begin{pmatrix}
\frac{c_1}{b+s} & 0 \\ 
0 & \frac{c_2 }{b+s^{-1}}
\end{pmatrix} \cdot \vec{r}_{\text{out}},
\end{split}
\end{equation}
as previously stated we will focus on the case $\vec{r}_\text{out} = (0,0)$, 
which is the most likely outcome according to \eqref{eq:p_rout}; this 
implies a conditional state with $\vec{r}'_{A} = (0,0)$.
\subsection{Squeezed thermal states}
We first focus on two-mode squeezed thermal states (STS), which means 
setting $c_1 = -c_2 = c$. We can get an intuition already by looking 
at the CM,
\begin{equation}
\begin{pmatrix}
a - \frac{c^2}{b+s} & 0 \\ 
0 & a - \frac{c^2}{b+s^{-1}}
\end{pmatrix};
\end{equation}
we see that for heterodyne measurements ($s=1$) we have a thermal state, 
which is incoherent, while the maximally squeezed state is obtained 
for homodyne measurements ($s\to \infty$).
\par
For STS we will just focus on squeezing along the $x$ direction since 
the direction of squeezing does not play a role, given the symmetry 
of the state. It can be useful to parametrize the measurement covariance 
matrix as $s = e^{2 r_m}$ and $1/s = e^{-2 r_m}$, where $r_m \geq 0$ is 
the ``physical'' squeezing parameter of the measurement; in the limit 
$r_m \to \infty$ ($s \to \infty$) we get an homodyne measurement of 
the quadrature $\hat{p}$.
\par
It is interesting to notice that conditional state~\eqref{eq:cond_state2modes} 
is insensitive to the sign of $c_1$ and $c_2$. The same results are obtained 
for states with $c_1 = c_2 = c$, which will dub in the following mixed thermal 
states (MTS), which are always separable and physically correspond to thermal 
states mixed with a beam splitter. It follows that the same remote coherence 
can be created from these two classes of states for fixed $a$, $b$ and $c$, 
however, the range of physically allowed values of $c$ is different in the 
two cases and STS can be more correlated, and in fact also entangled.
\par
In what follows we will consider both the regular and the Gaussian relative 
entropy of coherence $C_S$ and $C_S^G$. The measure $C_S$ is computed numerically by truncating the Fock space and by evaluating the corresponding
photon number distribution $\left\{ \langle n | \rho | n \rangle \right\}$ 
needed to evaluate the Shannon entropy for generic single mode Gaussian 
states~\cite{Marian1993a,Dodonov1994}.
\subsubsection{Symmetric STS}
As a first step, we focus on symmetric STSs, for which local thermal
states have the same energy, i.e., $a=b>1$. The parameter $c$ embodies the 
total correlations between the subsystems; for all these states, in order 
to  satisfy the physicality condition (\ref{eq:physical}), one needs 
$|c| \leq \sqrt{a^2 -1}$. We refer to the equality as to the 
physicality threshold, which is achieved by pure STS, i.e., the 
so-called {\em twin beam} states. 
On the other hand, separable states must satisfy the condition 
$|c| \leq a - 1$, also referred to as the separability threshold, 
corresponding to the physicality threshold for symmetric 
STSs with the same parameter $a$. 
We also employ the physical parametrization of STSs: $a=b=\left(1 + 2 N \right) 
\cosh 2r$ and $c_1= - c_2 =\left( 1 + 2 N \right) \sinh 2r$, 
where $N \geq 0 $ and $r\geq 0$ represent the number of thermal 
photons and a real squeezing parameter respectively. 
\begin{figure}[h!]
	\centering
	\includegraphics[height=5cm]{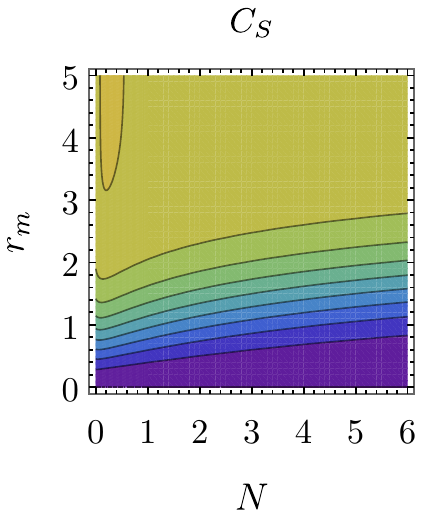}
	\includegraphics[height=5cm]{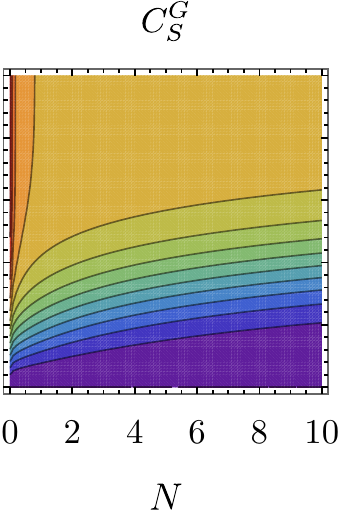}
	\includegraphics[width=.8cm,trim= .7cm -5.5cm 0cm 0cm,clip]{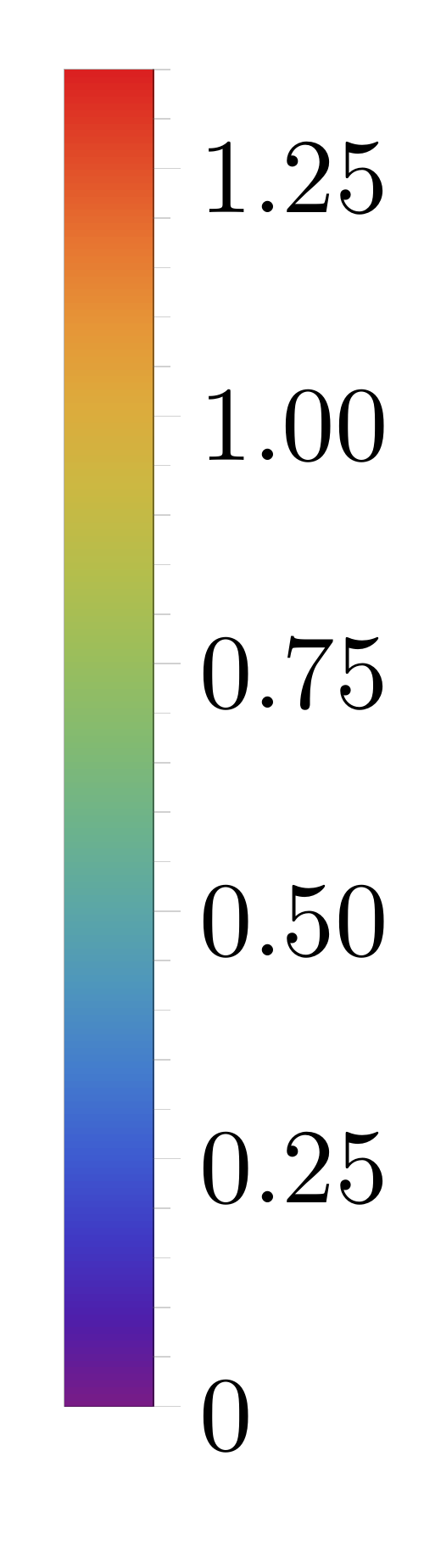}
	\caption{Relative entropy of coherence $C_S$ (left panel) and Gaussian 
	relative entropy of coherence $C_S^G$ (right panel) for the conditional state after Gaussian measurement
	of one mode of a symmetric STS. Coherence is shown as a function of the STS 
	thermal photons $N$ and the measurement squeezing $r_m$ for a fixed value 
	of the STS squeezing parameter $r=1$.}
	\label{fig:coh_N_rm}
\end{figure}
\par
In Fig. \ref{fig:coh_N_rm} we show the behavior of the relative 
entropy of coherence $C_S$ of the most probable conditional state 
and of its Gaussian version $C_S^G$ as a function of the number 
of thermal photons $N$ and the squeezing 
of the measurement $r_m$ at fixed initial squeezing $r$.
The behavior of both measures is similar: they both increase by increasing 
the squeezing of the measurement $r_m$ and reach an asymptotic value 
for homodyne measurements ($r_m \to \infty$). They are both decreasing functions 
of the number of thermal photons $N$ (at least for a sufficiently high $N$) 
and they tend to an asymptotic value dependent on $r_m$, as reported in~\cite{Buono2016}. 
The only relevant difference is that $C_S$ initially 
shows a slight increment as a function of $N$. We also correctly show 
that $C_S^G$ is an upper bound to $C_S$.
These results indeed support the idea that by projecting subsystem $B$ 
on a squeezed state we can generate coherence in subsystem $A$, even if 
the initial state is highly mixed. We have strong numerical evidence 
that for this class of states the remote coherence is a monotonous function 
of the squeezing of the measurement $r_m$ and thus that
homodyne measurement is optimal. This is in agreement with physical 
intuition since an homodyne detection (with zero outcomes) amounts to a 
projection on an infinitely squeezed vacuum state and this kind of state 
becomes more and more coherent as far as the squeezing increases.
\par
\begin{figure}[h!]
	\centering
	\includegraphics[width=8.4cm]{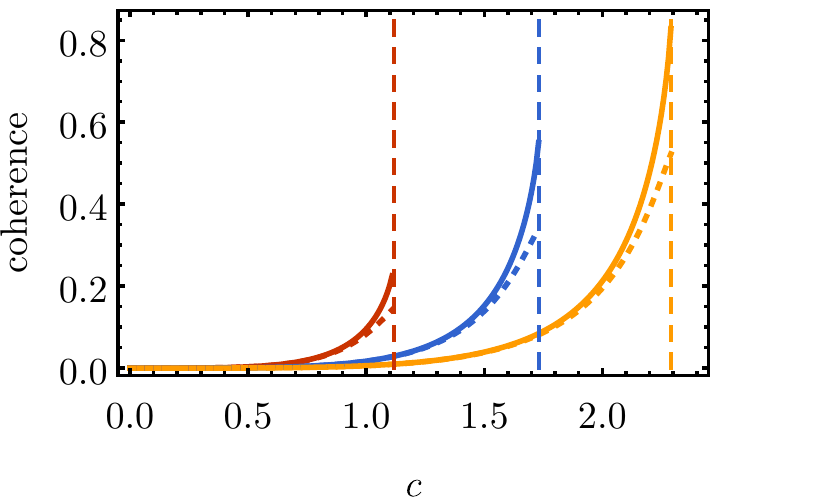}
	\caption{Gaussian relative entropy of coherence $C_S^G$ 
	(upper solid curves) and relative entropy of coherence $C_S$ 
	(lower dotted lines) as a function of the covariance matrix term $c$ 
	for a symmetric STS and homodyne measurement ($s\to \infty$). From 
	left to right the set of curves represents $a=1.5,2,2.5$ respectively 
	(red, blue and yellow), the vertical lines are drawn at the physicality
	 bound $c=\sqrt{a^2 -1}$.}
	\label{fig:coh_vs_c}
\end{figure}

In Fig.~\ref{fig:coh_vs_c} we show the maximal remote conditional coherence, 
obtained with an homodyne measurement, as a function of the parameter 
$c$, for different values of $a$, which also fixes the total energy 
of the state. We also have strong evidence that remote coherence is 
monotonically increasing in $c$ at fixed $a$, i.e. by increasing the 
two-mode squeezing at a fixed energy.
\par
Concerning the monotonicity
(as a function of $c$ and $r_m$) of the Gaussian measure $C_S^G$ we
can prove that the difference between the energy of the corresponding 
thermal state and the square root of the determinant of the covariance 
matrix (see Eq.~\eqref{eq:coh_rel_ent2b}) is a monotonically increasing 
function. However, $h(x)$ being a concave function, this does not imply 
the monotonicity of $C_S^G$ (but it is actually a condition implied by 
it). Overall, this suggests that our numerical result does indeed hold in 
general.
\par
\begin{figure}[h!]
	\centering
	\includegraphics[width=4.25cm]{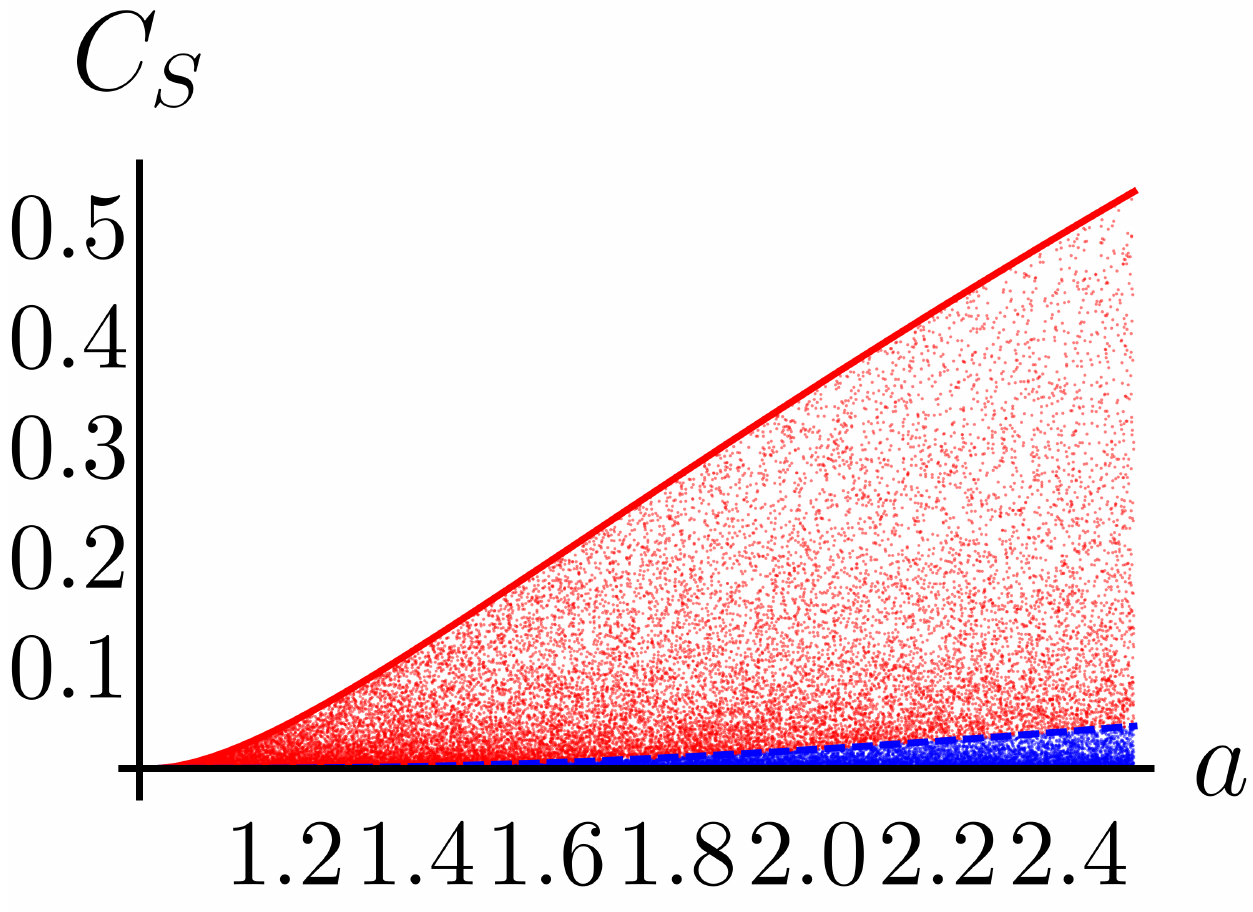}
	\includegraphics[width=4.25cm]{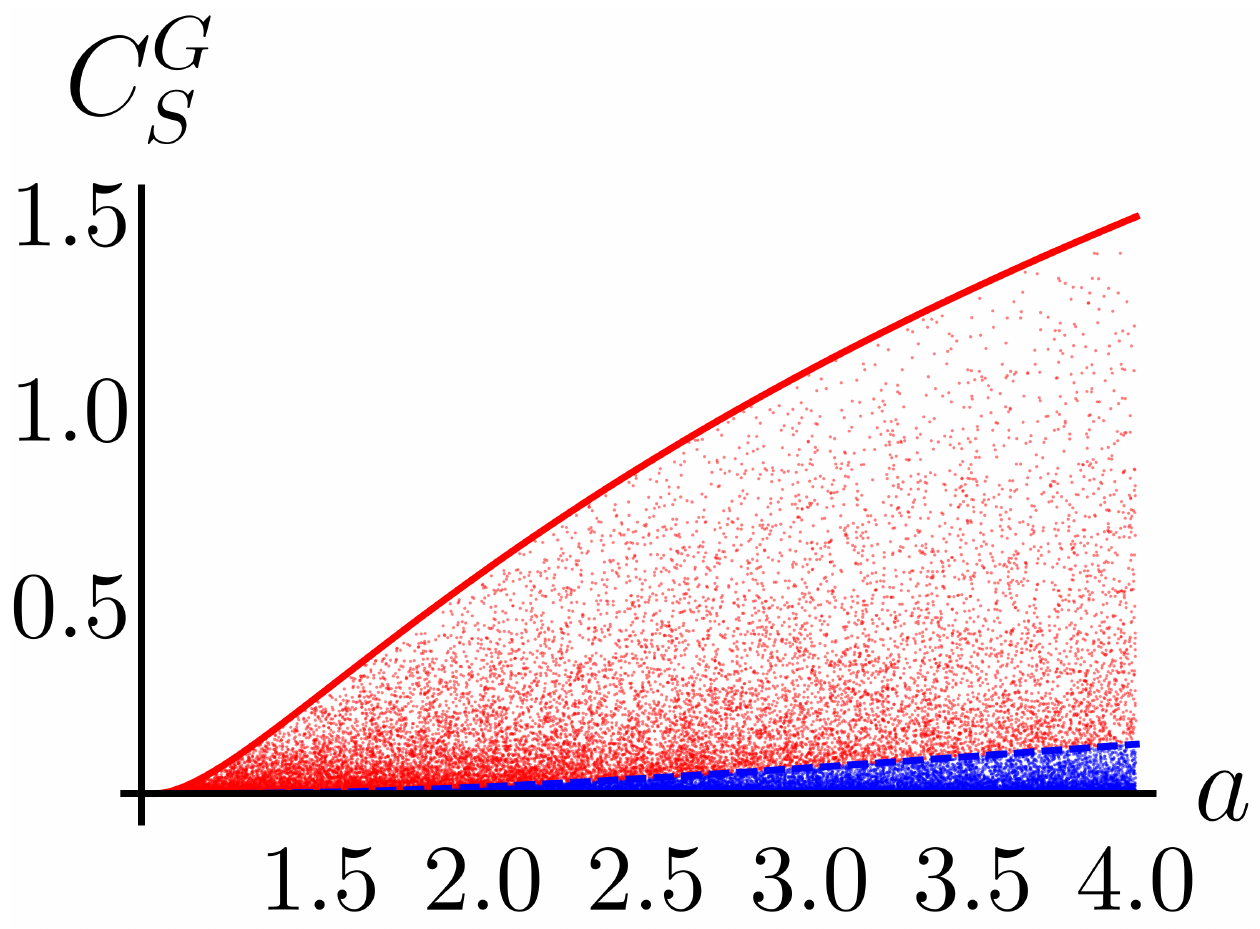}
	\caption{Relative entropy of coherence (left) and Gaussian relative 
	entropy of coherence (right) for homodyne measurement on mode $B$ of 
	symmetric STS, as a function of $a$. The $5\cdot 10^4$ random symmetric STS are generated 
	by sampling uniformly the parameters $a$ and $c$.
	The solid red (dashed blue) curve at the top (in the middle) represents the physicality (separability) threshold.
	Entangled (separable) states correspond to red (blue) points above (below) the separability threshold.}
	\label{fig:sep_symSTS_Gauss}
\end{figure}

The monotonic behavior of remote coherence as a function of $c$ implies 
that we can use this figure of merit for entanglement detection. Given 
the local energies $a=b$, there is a threshold value for the remote 
coherence which separates entangled and separable states.
A very similar behavior was observed in~\cite{Brunelli2017} considering the extractable work via Gaussian measurements as a figure of merit. A similar 
bound on separable states also arises by considering quantum discord~\cite{Adesso2010,Giorda2010}, in that case however also an 
energy-independent bound exists.
This feature is illustrated in Fig.~\ref{fig:sep_symSTS_Gauss}, where 
one can see that the remote coherence for randomly generated symmetric 
STS lie above the curve given at the separability threshold if and only 
if they are entangled, for both coherence measures.
\subsubsection{Asymmetric STS}
We now consider asymmetric STSs, with two distinct local energies $a>1$ and 
$b>1$. The physicality threshold is represented by the condition $|c| \leq 
\sqrt{ a b - 1 - | a - b|}$ while the separability threshold is set by the 
condition $ | c | \leq \sqrt{ a b +  1 - a - b}$, which corresponds to the physicality threshold for asymmetric MTS with the same parameters.
\begin{figure}[h!]
	\centering
	\includegraphics[width=4.2cm]{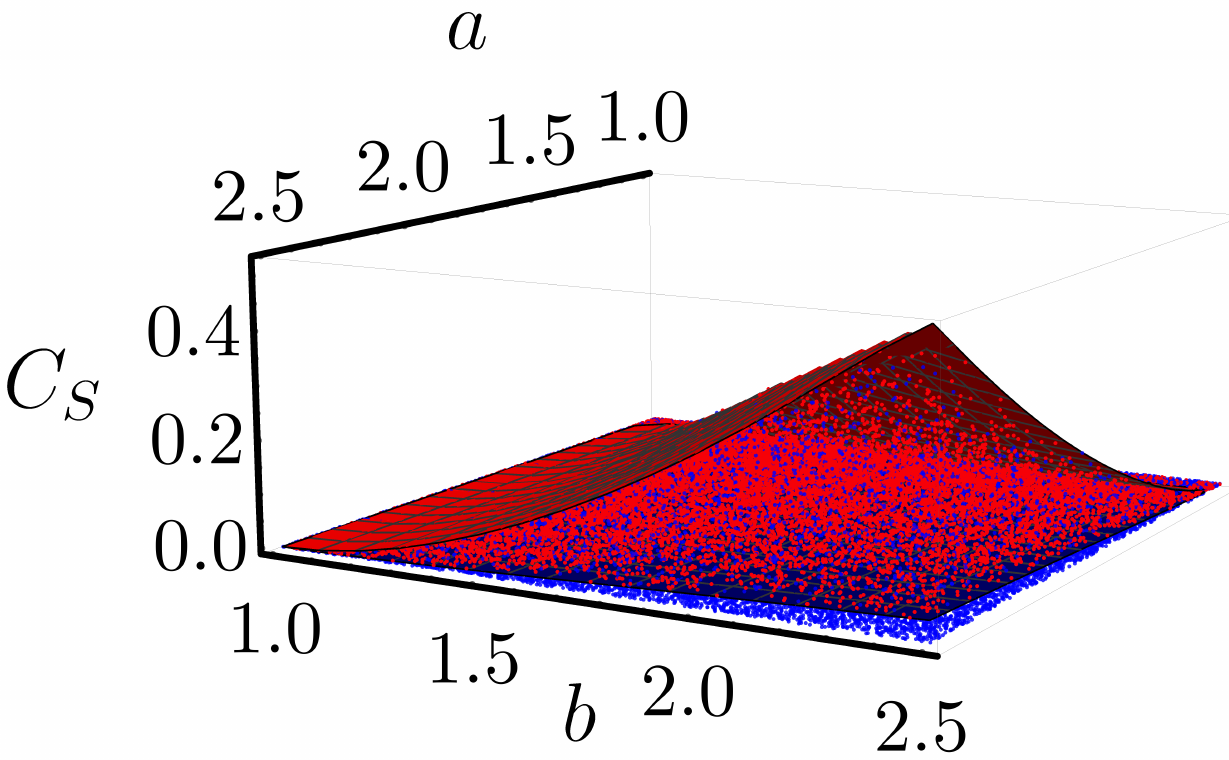}
	\includegraphics[width=4.2cm]{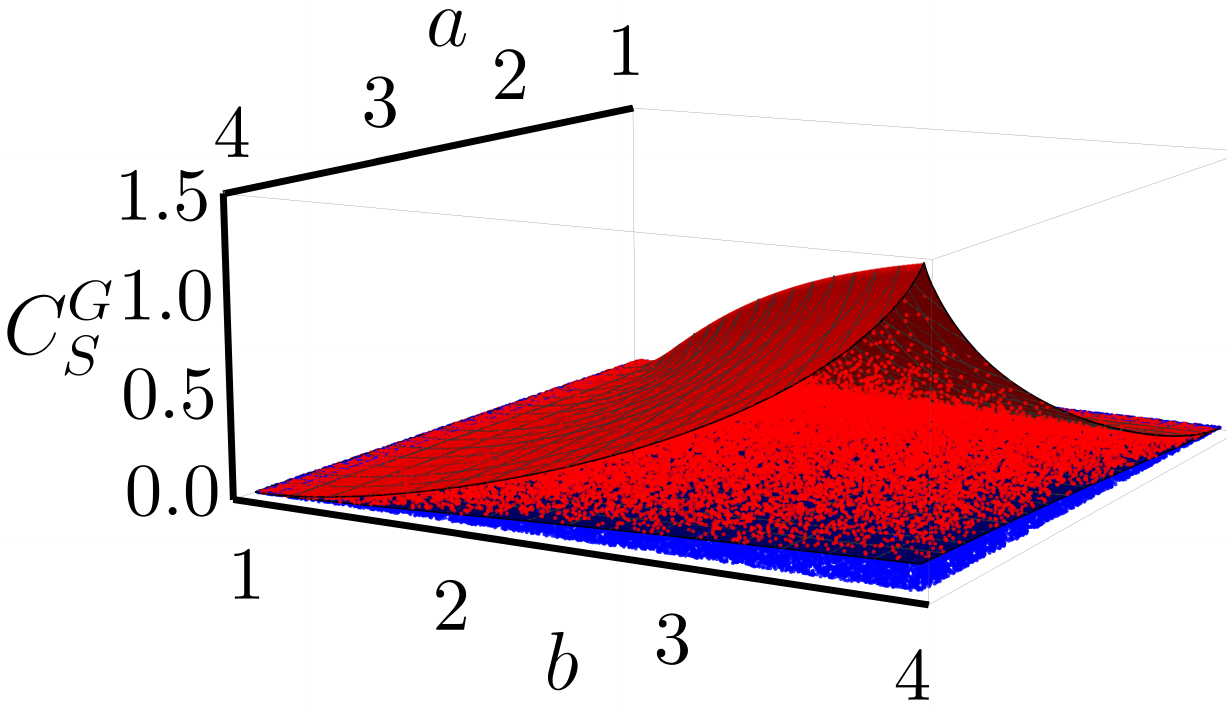}
	\caption{Relative entropy of coherence (left) and Gaussian relative 
	entropy of coherence (right) for homodyne measurements on mode $B$ of 
	asymmetric STS, as a function of the parameters $a$ and $b$. 
	The $5\cdot 10^4$ random asymmetric STS are generated by sampling 
	uniformly the parameters $a$, $b$ and $c$.
	The upper red (lower blue) surface corresponds to states on the phisicality (separability) threshold.
	Entangled (separable) states correspond to red (blue) points above (below) the separability threshold.}
	\label{fig:sep_asymSTS_Gauss}
\end{figure}
\par
Most of what we have learned for symmetric STSs still holds. We have numerical evidence that an homodyne measurement is optimal to remotely generate coherence and that remote coherence is a monotonically increasing function of $c$ and of $r_m$ at fixed $a$ and $b$. This means that we have a bound on the remote coherence which enables us to discriminate between separable and entangled states at fixed local energies, in complete analogy with the previous case. This is presented in Fig.~\ref{fig:sep_asymSTS_Gauss}, where we show that the remote coherence for randomly generated STSs lies above the surface at the separability threshold if and only if the states are entangled. The same results hold for both coherence monotones.
\subsection{Generic states in normal form}
We now consider the full class of standard form two-mode states ($c_1\neq \pm c_2$), the physicality and separability conditions are more involved and we do not report them explicitly (see~\cite{Pirandola2009} for a thorough analysis).
We, again, have numerical evidence that homodyning is optimal for remote generation of coherence. However a measurement of the quadrature $x$ or $p$ is optimal depending on which canonical variables are more correlated, i.e., whether $|c_1| > |c_2|$ or the opposite. In the following we focus on the optimized remote coherence, generated by homodyning the appropriate quadrature. 
\par
The covariance matrix of the conditional state after homodyne measurements is a function of only one of the parameters $c_1$ and $c_2$, depending on which quadrature is measured. The optimized remote coherence is thus a function of a single parameter $c_{\text{max}}=\max \left( |c_1| , |c_2| \right)$ and so the conjecture of monotonicity presented earlier still applies. At variance with STSs the entanglement of generic states in normal form is not a monotonic function of $c_{\text{max}}$ and one can find separable states with a greater $c_{\text{max}}$ than some entangled states.
This implies that remote coherence cannot be used for discriminating entangled and separable states in this class, but we still have a conjectured bound at fixed local purities. We have numerical evidence that the upper value of $c_\text{max}$ for separable states (and therefore the maximal remote coherence) is obtained for the class of states with $c_2=0$ and $|c_1| = \sqrt{\left( a^2 - 1\right)\left( b^2 -1 \right)/ (a b)}$, which are separable states at the physicality threshold. Obviously the roles of $c_1$ and $c_2$ could be exchanged.
\begin{figure}[h!]
	\centering
	\includegraphics[width=4.2cm]{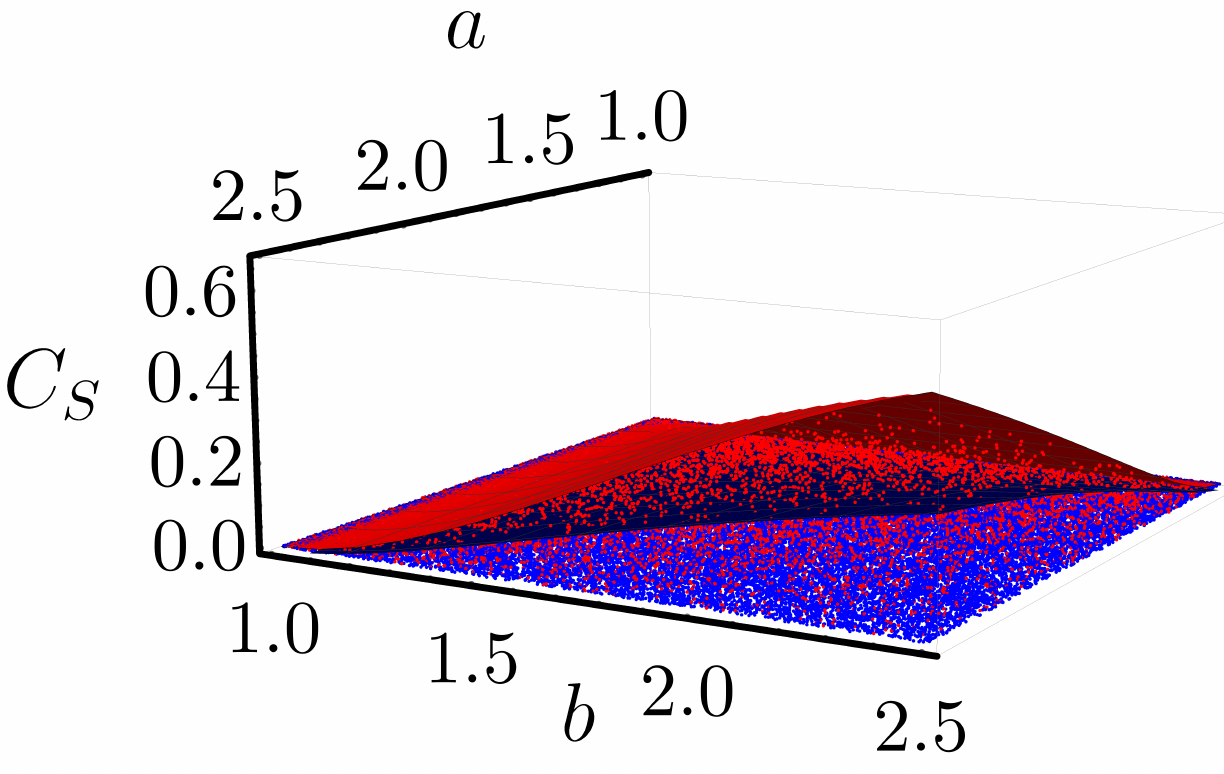}
	\includegraphics[width=4.2cm]{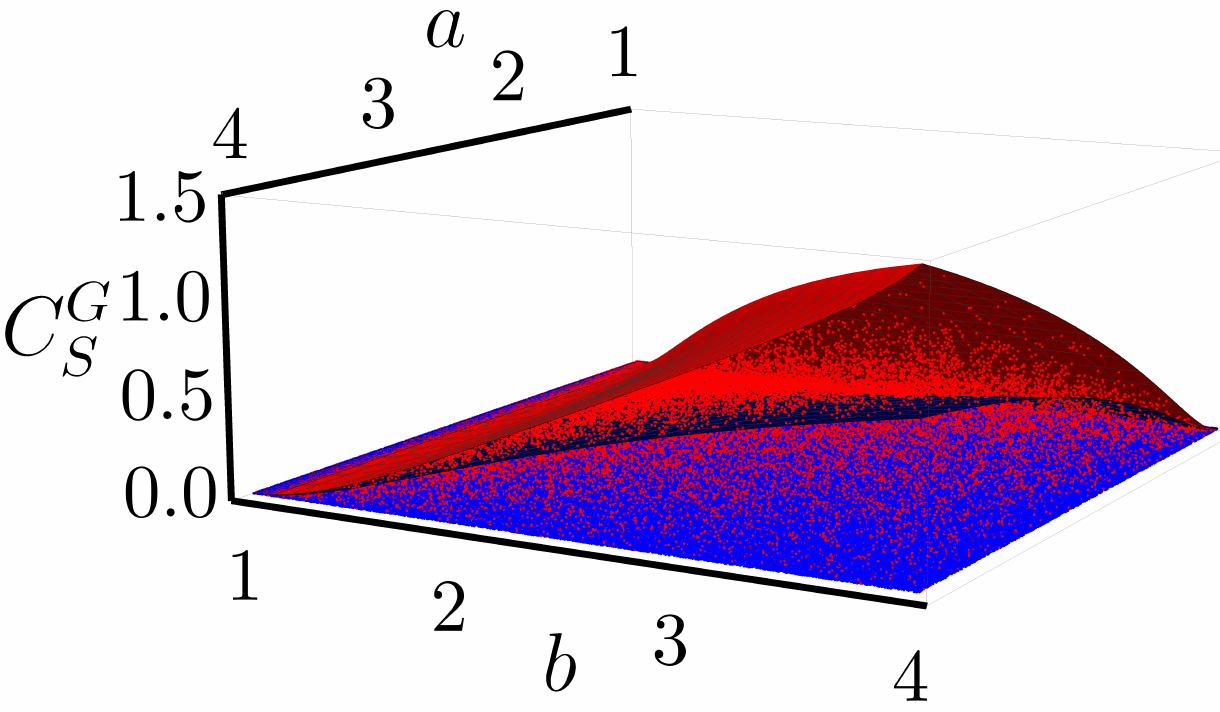}
	\caption{Relative entropy of coherence (left) and Gaussian relative entropy of coherence (right) for optimal homodyne measurements on mode $B$ of normal form 
	two-mode Gaussian state, as a function of the parameters $a$ and $b$. The $5\cdot 10^4$ (left) and $5\cdot 10^5$ (right) random states in normal form are generated by sampling uniformly the parameters $a$, $b$, $c_1$ and $c_2$.
	The upper red surface corresponds to states on the physicality threshold with maximal $|c_1|$, while the lower blue surface corresponds to states on the physicality the threshold with $c_2 = 0$.
	Entangled (separable) states correspond to ligher red (darker blue) points; only entangled states are above the lower blue surface.}
	\label{fig:sep_Generic_Gauss}
\end{figure}
\par
In Fig.~\ref{fig:sep_Generic_Gauss} we show the optimal remote coherence for random states in normal form; only entangled states lie above the surface given by separable states with maximal $c_{\text{max}}$. The upper surface is obtained by numerically maximizing $|c_1|$ at the physicality threshold for given $a$ and $b$ and coincides with pure STS states for $a=b$.	
These results also show that coherence due to measurement back-action can be stronger for separable states than for entangled ones. This is somewhat similar to what happens in the task of remote state preparation, where discordant resource states can outperform entangled states~\cite{Dakic2012}. However, in the present problem quantum discord is not a monotonic function of the remote coherence, therefore it cannot be regarded as a proper resource for the task.
\subsection{Feasible three-mode state}
In order to show that Gaussian measurements on a single mode can generate coherence and correlations in the bipartite conditional state we focus on a particular example: the pure tripartite obtained by interlinked bilinear interactions~\cite{Andrews1970,Ferraro2004,OLIVARES2013}, which is feasible experimentally.
The first moments of this state are null, while its CM is
\begin{equation}
\label{eq:interlinked_covM}
\Sigma_{\text{T}}=\begin{pmatrix}
\sigma_A & \sigma_{AB} & \sigma_{AC}  \\ 
\sigma_{AB} & \sigma_B  & \sigma_{BC} \\
\sigma_{AC} & \sigma_{BC} & \sigma_{C}
 \end{pmatrix},
\end{equation}
where 
\begin{equation}
\begin{split}
&\sigma_A = \left( 2 N_A + 1 \right) \mathbb{1} \quad \sigma_B = \left( 2 N_B + 1 \right) \mathbb{1} \quad \sigma_C = \left( 2 N_B + 1 \right) \mathbb{1} \\
&\sigma_{AB} = 2 \sqrt{N_B \left( N_A + 1 \right)  } \mathbb{P} \quad \sigma_{AC} = 2 \sqrt{N_C \left( N_A + 1 \right)  } \mathbb{P} \\
&\sigma_{BC} =  2 \sqrt{ N_B N_C} \mathbb{P}
\end{split}
\end{equation}
with $N_A = N_B + N_C$ and $\mathbb{P}=\text{diag}\left(1,-1\right)$. The expansion on the Fock basis is the following
\begin{equation}
\label{eq:interlinked_Fock}
\begin{split}
|\xi \rangle =& \frac{1}{\sqrt{1 + N_A}} \sum_{p,q} \left( \frac{N_B}{1+N_A} \right)^{p/2} \left( \frac{N_C}{1+N_A} \right)^{q/2} \\
 & \times  \left[ \frac{ \left( p + q \right)!}{p! q!} \right]^{1/2} | p + q, p, q \rangle.
\end{split}
\end{equation}
We discuss the situation where a Gaussian measurement is performed on mode $A$ and we study coherence and correlations of the conditional two-mode state of modes $B$ and $C$. In this case, the marginal state is not locally thermal, but it is correlated and has quantum coherence. 
\par
\begin{figure}[h!]
	\centering
	\includegraphics[width=4.25cm]{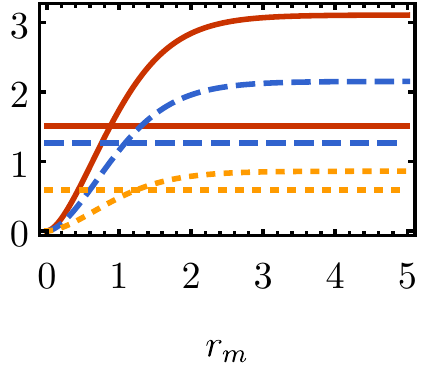}
	\includegraphics[width=4.25cm]{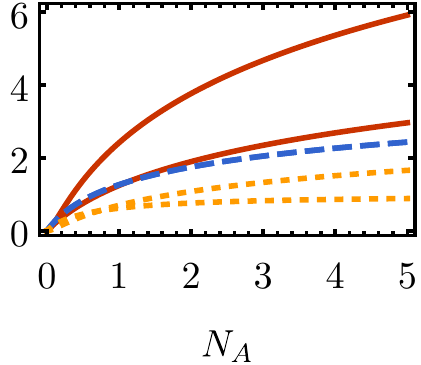}
	\caption{Coherence and discord of the two-mode conditional state
	obtained by measuring mode $A$ of the feasible three-mode state \eqref{eq:interlinked_Fock}. 
	In both panels the solid red curves represent $C_S^G$, 
	the dashed blue curves $C_S$ and the dotted orange curves the 
	quantum discord $D(B \!: \! C)$.
	In the left panel, the quantities are shown as a function of the measurement 
	squeezing $r_m$ at a fixed $N_B=1$ and $N_C =2$; the horizontal lines 
	correspond to the same figures of merit computed for the marginal state (and are thus independent from measurement squeezing). 
	In the right panel, the quantities are shown as a function of $N_A$ by fixing $N_B = N_C=N_A / 2$, with $r_m = 5$; 
	the lower curves correspond to the same figures of merit computed for the marginal state.}	
	\label{fig:Interl_Coh_disc}
\end{figure}

In Fig.~\ref{fig:Interl_Coh_disc} we show the measurement-induced coherence, using both coherence measures, and the measurement-induced quantum discord, as a function of the total energy of the state and of the measurement squeezing. Since the conditional state remains pure, in this case quantum discord reduces to entanglement entropy. We also report the coherence and discord of the marginal state, explicitly showing that if measurement squeezing is not high enough we obtain values lower than the ones we obtain by studying the initial marginal states. Furthermore, we correctly see that, as predicted by inequality (\ref{eq:bounddiscord}), coherence of the two-mode state is always an upper bound to discord.
\section{Coherence via Continuous Gaussian Measurements}
\label{sec:continuous}
We now consider a different protocol for producing single mode coherence,
based on continuous monitoring of the environmental degrees of freedom via Gaussian measurements. This setting bears some similarities to the protocol for remote creation of coherence considered in the previous Section. In this case, the 
necessary correlations between the environment and the system are provided 
by the dynamics.
\subsection{Gaussian conditional dynamics}
We briefly review the notation and formalism needed to describe Gaussian 
conditional dynamics (see~\cite{Genoni2016} for further details). We deal 
with a bosonic system in a Gaussian state, described by a covariance matrix 
$\sigma$ and first moment vector $\vec{r}$. At each instant of time, the system interacts with a Markovian bath, described by input operators $\vec{r}_\text{in}(t)$ and correlation matrix $\sigma_E$, via a bilinear Hamiltonian
\begin{equation}
\hat{H}_C = \vec{\hat{r}}^{\intercal} C \vec{\hat{r}}_\text{in}(t),
\end{equation}
where $C$ is an arbitrary matrix. If we trace out the degrees of freedom of the 
bath, i.e. we do not record measurements on the environmental degrees of freedom, 
the dynamics of the CM is described by a diffusion equation,
\begin{equation}
\label{eq:diffusion}
\dot{\sigma}= A \vec{\sigma} + \vec{\sigma} A^{\intercal} + D,
\end{equation}
where 
\begin{equation}
A = \Omega H_s + \frac{\Omega C \Omega C^{\intercal}}{2}, \quad D = 
\Omega C \sigma_E C^{\intercal} \Omega^{\intercal}.
\end{equation}
If we also assume that the bath has zero first moments and that the system 
is not driven, the differential equation for the first moments then reads
\begin{equation}
\label{eq:unmon_first_moments}
\vec{\dot{r}}'=A \vec{r}'\,.
\end{equation}
Gaussian states are completely defined by first and second moments and thus 
one may derive the standard master equation in Lindblad form for the density
 operator from Eqs. \eqref{eq:diffusion} and \eqref{eq:unmon_first_moments}.
 \par
If we introduce the continuous monitoring of the environmental modes through a Gaussian measurement described by a matrix $\sigma_m$, we find that the CM obeys a deterministic Riccati equation
\begin{equation}
\label{eq:riccati}
\dot{\sigma'} = \tilde{A} \sigma' + \sigma' \tilde{A}^{\intercal} + \tilde{D} - \sigma' B B^{\intercal} \sigma',
\end{equation}
where we have defined
\begin{align}
\label{eq:tilde_matr}
\tilde{A} &= A - \Omega C \sigma_E \left( \sigma_E + \sigma_m \right)^{-1} \Omega  C^{\intercal} \\
\tilde{D} & = D + \Omega C \sigma_E \left( \sigma_E + \sigma_m \right)^{-1} \sigma_E C^{\intercal} \Omega \\
B &= C \Omega \left( \sigma_E + \sigma_m \right)^{-\frac{1}{2}};
\end{align}
as in the previous section, the CM $\sigma_M$ defines a generic Gaussian measurement.
On the contrary, the first moments conditional evolution is stochastic and governed by
\begin{equation}
\label{eq:r_wiener}
\mathrm{d} \vec{r'} = A \vec{r'} \mathrm{d} t + \left( \Omega C \sigma_E - \sigma'	 C \Omega \right) \left( \sigma_E + \sigma_m \right)^{\frac{1}{2}} \frac{\mathrm{d}\vec{w}}{\sqrt{2}},
\end{equation}
which is an Ito stochastic differential equation corresponding 
to a  classical Wiener process; the vector of Wiener increments
 $\mathrm{d}\vec{w}$ satisfies $\mathrm{d} w^2_j = \mathrm{d} t$.
\subsection{Coherence of a monitored quantum optical parametric oscillator}
We now focus on the simple model of a single mode quantum optical parametric oscillator, physically composed by an optical cavity mode driven by a pump 
laser and interacting with a nonlinear optical crystal. The effective 
Hamiltonian for the system is
\begin{equation}
\label{eq:opo_ham}
	H_s=-\frac{\chi}{2} \left( \hat{x} \hat{p} + \hat{p} \hat{x}\right),
\end{equation}
where $\hat{x}$ and $\hat{p}$ are conjugated quadratures of the field
being amplified and $\chi$ is a coupling constant given by the second 
order nonlinear coefficient of the crystal times the average photon 
number of the pump laser. We consider the system interacting with a 
Markovian bath at thermal equilibrium, which can be described by a 
single mode CM of the form
\begin{equation}
\sigma_E = \frac1\mu \mathbb{1}_2 \quad
\mu = \left( 2 N +1 \right)^{-1}\,.
\end{equation}
The interaction between the cavity mode and the environment is passive and
modeled by the Hamiltonian
\begin{equation}
	H_C  = \sqrt{\gamma} \left( \hat{x} \hat{x}_{in}(t) + \hat{p} \hat{p}_{in}(t) \right),
\end{equation} 
such that the coupling matrix reads $C=\sqrt{\gamma} \mathbb{1}_2$.
If the environment is left unmonitored then the unconditional dynamics 
is described by the standard quantum optical master equation. 
The unconditional dynamics is stable and admits a steady state for $\chi < \frac{\gamma}{2}$, the steady state CM is found by imposing $\dot{\sigma}=0$ 
in Eq.~\eqref{eq:diffusion} and reads
\begin{equation}
\label{eq:uncond_ss}
\sigma_{\text{ss}} = \frac1\mu\, \begin{pmatrix}
 \frac{1}{1 + 2 \tilde{\chi}} & 0 \\
0 & \frac{1}{1 - 2 \tilde{\chi}}
\end{pmatrix},
\end{equation}
while the first moments are null, for convenience we have defined $ \tilde{\chi}= \frac{\chi}{\gamma}$, so that the stability condition becomes $\tilde{\chi} \leq \frac{1}{2}$.
\par
The steady state is clearly squeezed and thus has nonzero quantum coherence. 
We will now show that its quantum coherence can be improved thanks to environmental monitoring, if the measurements are projections on states which are squeezed enough. For general-dyne monitoring with unit efficiency and real squeezing parameter, i.e. a pure and diagonal $\sigma_m$ as defined in the previous section, the matrices \eqref{eq:tilde_matr} become
\begin{align}
\label{eq:tilde_matr_opo}
\tilde{A} &= 
\begin{pmatrix}
-\chi - \frac12\frac{\gamma \mu e^{-2 r_m}}{1 + \mu e^{-2 r_m}} & 0 \\ 
0 & \chi - \frac12\frac{\gamma \mu e^{2 r_m} }{1 + \mu e^{2 r_m}} 
\end{pmatrix} \\
\tilde{D} & = \frac{\gamma}{\mu}\,\begin{pmatrix}
\frac{ \mu e^{-2 r_m}}{1 + \mu e^{-2 r_m}} & 0 \\
0 & \frac{\mu e^{2 r_m}}{1 + \mu e^{2 r_m}}
\end{pmatrix}\\
B &= \sqrt{\mu\gamma}\:\begin{pmatrix}
0 & \sqrt{\frac{1}{1 + \mu e^{-2r_m}}} \\
-\sqrt{\frac{1}{1 + \mu e^{2r_m}}} & 0.
\end{pmatrix}
\end{align}
The steady state CM of the conditional dynamics is obtained by setting $\dot{\sigma}=0$ in Eq.~\eqref{eq:riccati}, the result of the algebraic equation is the following
\begin{equation}
\label{eq:ssCMgd}
\sigma'_{\text{ss}} = \begin{pmatrix}
\frac{ \tilde{A}_{11} + \sqrt{ \tilde{A}_{11}^2 + (B B^\intercal)_{11} \tilde{D}_{11}}}{ (B B^\intercal)_{11}} & 0 \\
0 & \frac{ \tilde{A}_{22} + \sqrt{ \tilde{A}_{22}^2 +  (B B^\intercal)_{22} \tilde{D}_{22}}}{(B B^\intercal)_{22}} ,
\end{pmatrix}
\end{equation}
which again is a function of the parameter $\tilde{\chi}$. For homodyne detection of $\hat{p}$ ($r_m \to \infty$) we obtain a thermal squeezed state with exactly $N$ thermal photons and a squeezing parameter dependent on $\tilde{\chi}$, described by the following CM
\begin{equation}
\sigma_{\text{ss}}^{\text{\sf(hom)}} = \frac1\mu\,\begin{pmatrix}
1 - 2 \tilde{\chi}  & 0 \\
0 & \frac{1}{1 - 2 \tilde{\chi}} 
\end{pmatrix}.
\end{equation}
This scenario is similar to the one we have studied in the previous Sections. 
Also in this case, we neglect the first moments of the steady state. Indeed, 
the zero first moments case corresponds to the most likely event. In addition, 
the coherence achievable by nonzero first moments may be achieved by 
displacing the state afterward as well.
\par
\begin{figure}[h!!]
	\centering
	\includegraphics[height=5.2cm]{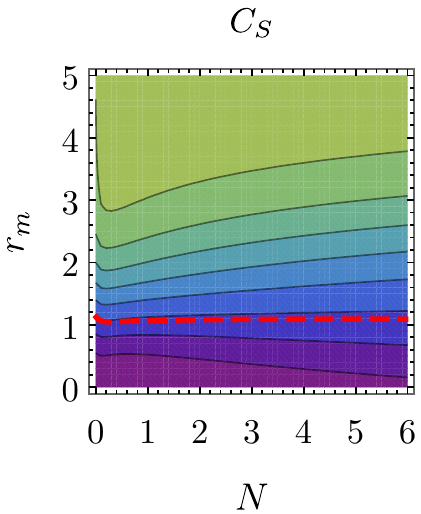}
	\includegraphics[height=5.2cm]{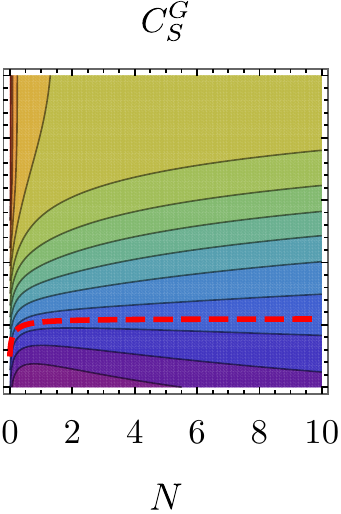}
	\includegraphics[width=.8cm,trim= .7cm -5.5cm 0cm 0cm,clip]{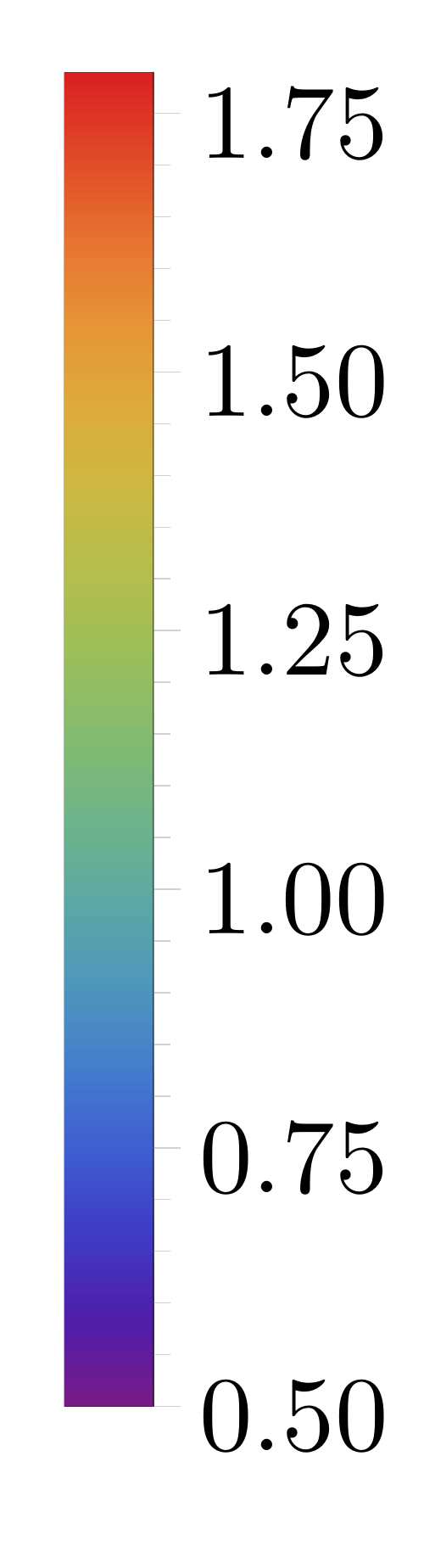}
	\caption{Steady state coherence (at zero first moments). 
	In the left panel we have the relative entropy of coherence and in the right panel its Gaussian counterpart, both as a function of the measurement squeezing $r_m$ and the mean number of excitation of the environmental mode $N$, with fixed $\tilde{\chi}=0.4$. The red dashed lines represent the threshold values $r_m^{\mathrm{th}}$ for which the coherence of the monitored state is equal to that of the unconditional dynamics. In the region above the red curve we have more coherence than for unconditional dynamics; in region below the curve, vice versa.}	
	\label{fig:ContCoh}
\end{figure}

In Fig.~\ref{fig:ContCoh} we show both measures of coherence for the monitored steady state as a function of the mean thermal excitations $N$ of the environmental 
state and of the measurement squeezing $r_m$. We find again that, neglecting first 
moments, the best possible measurement is homodyne detection, whereas a 
certain amount of squeezing is needed to surpass the coherence of the 
unmonitored state.
\par
\par
\begin{figure}[h!]
	\centering
	\includegraphics[width=8.4cm]{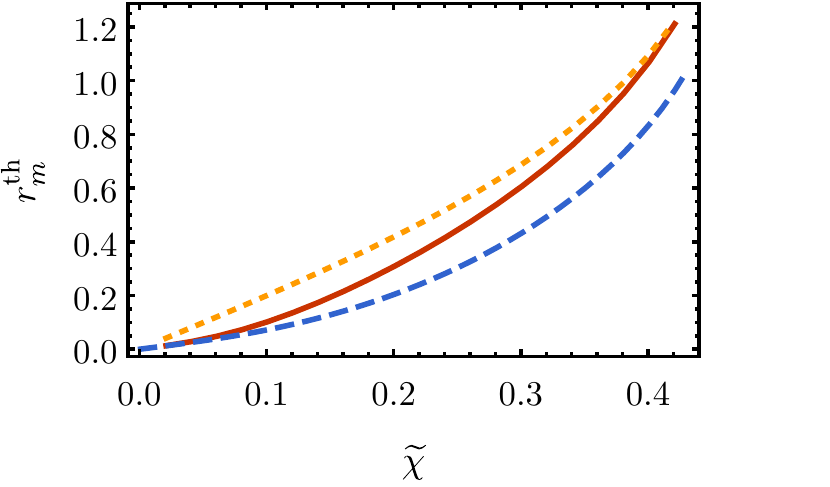}
	\caption{Value of the threshold squeezing as a function of $\tilde{\chi}$. The solid red curve is obtained for $C_S$ while the dashed blue for $C_S^G$, both for $N=0.1$; the dotted yellow curve represents both $C_S$ and $C_S^G$ for $N=5$.}	
	\label{fig:SqueezThresh}
\end{figure}

In Fig.~\ref{fig:SqueezThresh} we show the threshold value of the measurement squeezing $r_m^{\mathrm{th}}$ for which the coherence of the monitored state is equal to that of the unconditional dynamics (always neglecting first moments). We see that it is an increasing function of $\tilde{\chi}$; as a matter of fact for $\tilde{\chi} \to \infty$ the unconditional state becomes more and more squeezed, therefore an homodyne measurement is needed to achieve the same coherence in the conditional state.
In general, the two coherence monotones produce different threshold values, but this is noticeable only in the low $N$ regime and for larger $N$ the curves are indistinguishable. Moreover, in Fig.~\ref{fig:ContCoh} we also show $r_m^{\mathrm{th}}$ as a function of $N$ for a particular fixed value of $\tilde{\chi}$ and we see that it quickly saturates to an asymptotic value for growing $N$.
\par

\section{Conclusions}
\label{sec:conclusions}
In this paper, we have addressed the measurement-based generation of quantum 
coherence in continuous variable systems and investigated the coherence 
induced by Gaussian measurements on correlated Gaussian systems. 
\par
We have first explored a scenario for remote creation of coherence and 
analyzed in some detail the interplay with classical and quantum 
correlations. Starting from bipartite squeezed thermal states the remote 
coherence created by Gaussian measurements is a monotonic function of the 
relevant off-diagonal term of the covariance matrix, which in turn 
expresses the correlations among the 
canonical observables of the two parties. Given the symmetry 
of STSs, also entanglement and discord are monotonic functions of the same 
parameter. As a consequence, conditional coherence induced by measurement
may be used to discriminate between entangled and separable states, given 
the local purities. This is no longer true for the case of a generic two-mode 
state in normal form, for which we found a sufficient condition for detecting 
entangled states. A key finding is that measurement-induced coherence is 
not directly linked to quantum correlations, but rather to classical 
correlations between the two parties.
\par
We have also evaluated the conditional coherence achievable by conditional 
measurements on a specific class of three-mode states which are experimental feasible. From this we highlighted that measurement on a single mode induces 
both coherence and quantum correlations on the remaining two-mode system. In particular, we have shown that two-mode coherence on the Fock basis is an 
upper bound to the quantum discord.
\par
We then explored the coherence achievable by the continuous monitoring
of the environment of a continuous variable system. 
In particular, we have discussed the dynamics of an optical parametric 
oscillator and investigated how the coherence 
may be increased by means of time-continuous
Gaussian measurement on the interacting environment. 
In this case, we found that also the 
unconditional state has nonzero coherence, but there exists
a threshold on the measurement squeezing above which coherence is enhanced 
by the conditional measurement.
\par
Overall, our results show that Gaussian measurements represent a 
resource to create conditional coherence, which in turn may be exploited
as an entanglement witness.
\begin{acknowledgments}
This work has been supported by EU through the projects QuProCS 
(grant agreement 641277) and ConAQuMe (grant agreement 701154).
\end{acknowledgments}
\appendix
\section{Nonzero outcomes and average coherence}
\label{sec:appendix}
In this Appendix, we relax the assumption of zero measurement outcomes. 
We explore the effect of first moments on the remote creation of coherence 
and we also look at the average coherence w.r.t. the probability 
distribution of the outcomes. For simplicity, we restrict the analysis 
to two-mode STSs.
\subsection{First moments of the conditional state}
Let us consider a two-mode STSs with covariance matrix $\sigma$ and 
a general-dyne measurement, characterized by the CM
of a pure single mode state:
\begin{equation}
\begin{split}
\sigma_m &= 
R(\phi)\, \begin{pmatrix}
s  & 0 \\
0  & 1/s 
\end{pmatrix}\, R(\phi)^{\sf T} = \\ 
&= 
\begin{pmatrix}
 s \cos ^2 \phi +\frac{\sin ^2 \phi  }{s} & \frac{\left(s^2-1\right) \cos \phi  \sin \phi }{s} \\
 \frac{\left(s^2-1\right) \cos \phi  \sin \phi }{s} & \frac{\cos ^2 \phi }{s}+s \sin ^2 \phi  \\
\end{pmatrix};
\end{split}
\end{equation}
the covariance matrix $\sigma'_A$ and first moments $\vec{r}'_A$ of the conditional state on mode $A$ after the measurement $\sigma_m$ is performed on mode $B$ are given by Eqs.~\eqref{eq:cond_state}. 
We now want to explicitly evaluate the mean number of excitations due to the first moment of this state, i.e. $ \frac{1}{2} | \vec{r}'_A |^2$, as the Gaussian measure of coherence \eqref{eq:coh_rel_ent2b} monotonously depends on it.
We write the outcome of the measurement in polar coordinates $\vec{r}_{\text{out}} = ( |\vec{r}_{\text{out}}| \cos \theta,|\vec{r}_{\text{out}}| \sin \theta) $ and evaluate the term depending on the first moments explicitly: 
\begin{align}
\label{eq:first_momSTS}
\left| \vec{r}'_A \right|^2  = &\, \frac{ c^2 |\vec{r}_{\text{out}}|^2 s}{(b+s) 
(b s+1)} \Bigg\{ \left(b^2-1\right)\, +  \notag \\ 
& \left.   +\, \left(2 b+s+\frac{1}{s}\right) \left[
s \sin^2 (\theta-\phi)
+\frac{\cos^2 (\theta-\phi)}{s}
 \right] \right\}\,.
\end{align}
As it is apparent from the above formula, the relevant parameter is the 
relative angle $\phi - \theta$ between the squeezing and the measurement 
outcome vector.
Without loss of generality we can choose $s\geq1$; in this case the energy is maximized by $\phi - \theta = (k +1/2)\pi $, with $k \in \mathbb{Z}$.
\par
For heterodyne measurement ($s = 1$) the dependence on the angles is suppressed, resulting in
\begin{equation}
\left| \vec{r}^{\sf (het)} _A \right|^2 = \frac{c^2 |\vec{r}_{\text{out}}|^2}{(b+1)^2},
\end{equation}
while in the homodyne limit $s \to \infty$ we have 
\begin{equation}
\left| \vec{r}^{\sf (hom)} _A \right|^2 = \frac{c^2  |\vec{r}_{\text{out}}|^2 \sin ^2 \left(\theta - \phi \right)}{b^2}.
\end{equation}
\subsection{Remote coherence for non-zero outcomes}
\begin{figure}[h!]
	\centering
	\includegraphics[width=8.4cm]{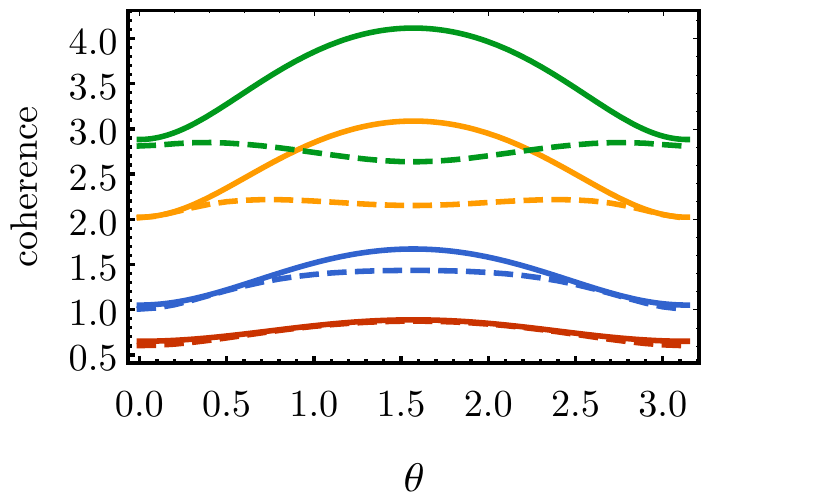}
	\caption{Remote coherence as a function of the outcome angle $\theta$ for symmetric STS with $N=1$ and $r=1$. From bottom to top we have $| \vec{r}_\text{out} | = 1,2,4,6$ in red, blue, yellow and green respectively. The dashed lines represent relative entropy of coherence $C_S$, while the solid ones the Gaussian counterpart $C_S^G$.}
	\label{fig:appendixTheta}
\end{figure}

A non-zero measurement outcome $| \vec{r}_{\text{out}} | \neq 0$ implies a non-zero first moments vector $\vec{r}'_A$ of the conditional state, which in turn means a higher coherence than in the zero outcome case. This is evident in Eq. \eqref{eq:coh_rel_ent2b} for the Gaussian relative entropy of coherence, but it is true also for the relative entropy of coherence.
As shown in the previous section, the quantity $\left| \vec{r}'_A \right|^2$ depends on the angle $\phi - \theta$. In particular if we fix the measurement angle $\phi = 0$ the maximum is for $\theta = \frac{\pi}{2}$, i.e. when the outcome vector is displaced along the $p$ axis. The same behavior is shared by the Gaussian measure of coherence, which is maximal for $\theta=\frac{\pi}{2}$ at fixed $|\vec{r}_{\text{out}}|$. This is shown in Fig.~\ref{fig:appendixTheta}, where we also show that the behavior of the relative entropy of coherence is different in general.
\par
\begin{figure}[h!]
	\centering
	\includegraphics[width=4.2cm]{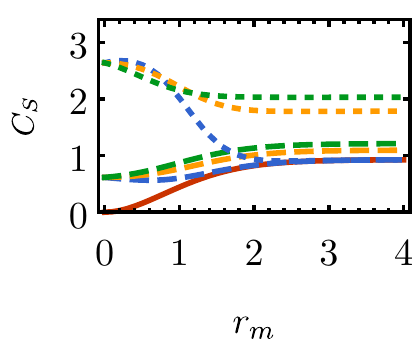}
	\includegraphics[width=4.2cm]{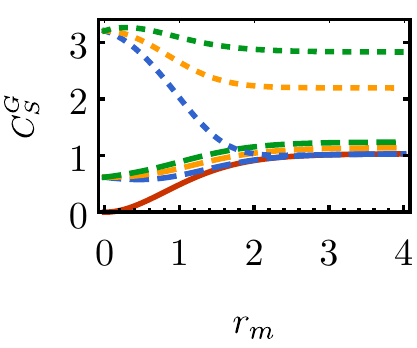}
	\caption{Relative entropy of coherence (upper panel) and Gaussian relative entropy of coherence (lower panel) as a function of $r_m$ for a symmetric STS with $N=1$ and $r=1$. The lowest solid red curve is for zero outcome $ | \vec{r}_\text{out} | = 0 $. The three dashed curves in the middle are for $ | \vec{r}_\text{out} | = 1$ and $\theta = 0,\frac{\pi}{4},\frac{\pi}{2}$ (from bottom to top), in blue, yellow and green respectively. The three dotted curves at the top are for $ | \vec{r}_\text{out} | = 4$ and $\theta = 0,\frac{\pi}{4},\frac{\pi}{2}$ (from bottom to top in the region $r_m \approx 2$), in blue, yellow and green respectively.}
	\label{fig:appendixFirstMom}
\end{figure}

In Fig. \ref{fig:appendixFirstMom} we show remote coherence as a function of $ r_m=\frac{1}{2} \log{s}$ for $\phi=0$ (as $ r_m \to \infty $ it becomes a measurement of $\hat{p}$ ) for different values of the outcome vector $\vec{r}_{\text{out}}$.
At variance with the case studied in the text, measurement squeezing can actually decrease the amount of coherence obtainable, depending on the value of the outcome vector. Moreover we can see again the different behavior of the two coherence measures, evident in the curves obtained for $| \vec{r}_\text{out} | = 4$.
These considerations resemble one of the results in~\cite{Ma2016a}, where the coherence generated by selective measurements is upper bounded by a term inversely proportional to the probability of getting the final state (calculation carried out for finite dimensions using the $l_1$ norm of coherence). 
In a similar way, in our Gaussian scenario the greater $|\vec{r}_{\text{out}}|$ is, the smaller the value of the probability density $p\left( \vec{r}_{\text{out}} \right)$. For fixed $|\vec{r}_{\text{out}}|$ and $\phi = 0$, the more displaced the state in the $p$ direction, the lower the value of the probability of getting that state. This happens because a Gaussian measurement with $\phi=0$ and $s>1$ has a Gaussian distribution of the outcomes $p\left( \vec{r}_{\text{out}} \right)$ which is ``squeezed'' along the $x$ axis.
\subsection{Average remote coherence}
\begin{figure}[h!]
	\centering
	\includegraphics[width=6cm]{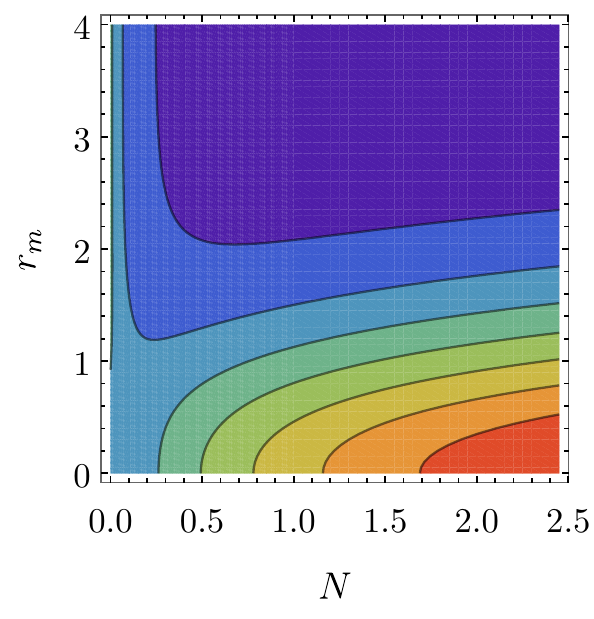}
	\includegraphics[width=.8cm,trim= .7cm -10cm 0cm 0cm,clip]{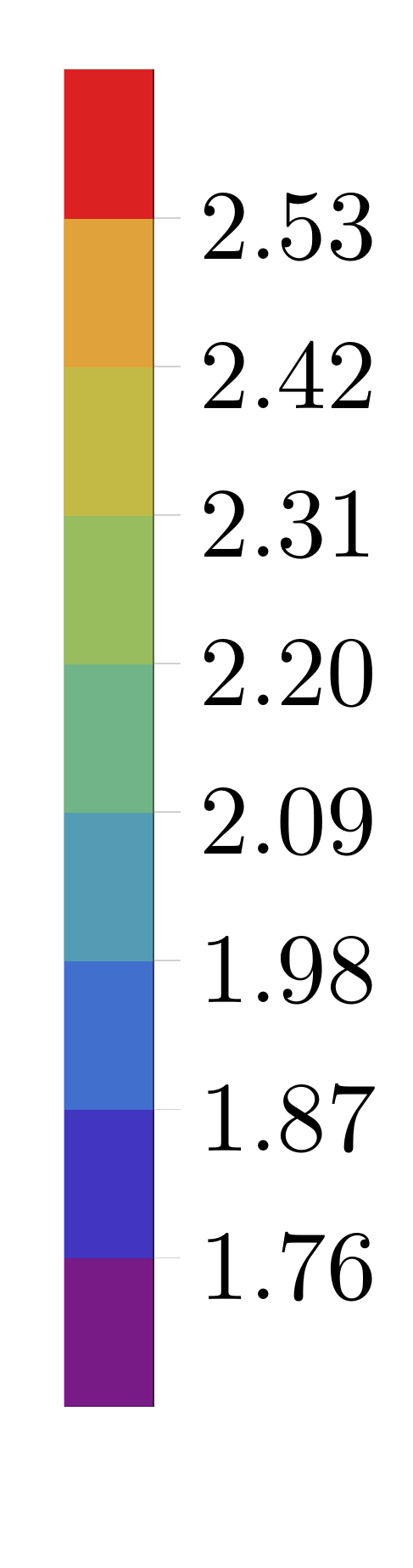}
	\caption{The average remote Gaussian relative entropy of coherence 
$\bar{C}^G_S$ for fixed $r = 1$ as a function of $r_m$ and $N$.}
	\label{fig:ave_coh1}
\end{figure}

\begin{figure}[h!]
	\centering
	\includegraphics[width=8.4cm]{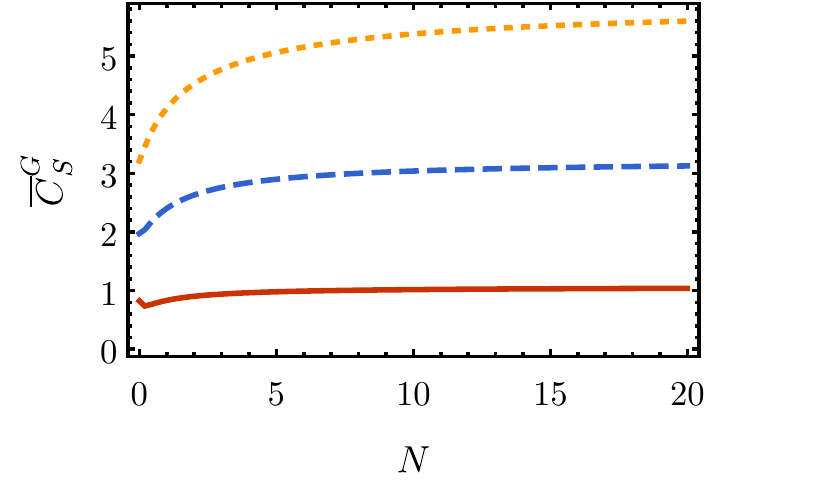}
	\caption{The average remote Gaussian relative entropy of coherence $\overline{C}^G_S$ for heterodyne measurements ($r_m = 0$) and symmetrical STS as a function of $N$. Different curves represent different values of the initial squeezing: $r = 0.5,1,1.5$ (full red, dashed blue and dotted yellow respectively).}
	\label{fig:ave_coh_het}	
\end{figure}

By dropping the zero outcome assumption, the most interesting quantity 
to consider is the average coherence that can be harvested if 
nonselective measurements are made on subsystem $B$ and all possible 
results are recorded. This figure of merit has been studied at length~\cite{Ma2016a,Hu2016a}. Given a coherence measure $C(\rho)$, in our 
continuous variable setting it is defined as 
\begin{equation}
	\overline{C}^{A | B} = \int \! \mathrm{d}^2 \vec{r}_\text{out} \, p(\vec{r}_\text{out}) {C}\left( \rho'_{A}\right),
\end{equation}
where $p(\vec{r}_\text{out})$ is the Gaussian distribution given by Equation \eqref{eq:p_rout}; in the following, we will omit the superscript $A | B$, because we always consider measurements on subsystem $B$ and we are interested in the coherence of system $A$. 
In order to compute $\bar{C}_S^{G}$ the integral has to be evaluated numerically; computing $\bar{C}_S$ is trickier because there is no closed formula for $C_S$ in the Gaussian case.
The contour plot of the average Gaussian relative entropy of coherence as a function of $N$ and $r_m$ for a symmetrical STS is shown in Fig.~\ref{fig:ave_coh1}.
We see that a heterodyne measurements yields the best average remote Gaussian coherence, at variance with the case of null outcomes. Moreover, the average Gaussian coherence actually increases with more mixed initial states, i.e. increasing $N$. This fact is explicitly shown in Fig.~\ref{fig:ave_coh_het}, where we report the results for heterodyne measurements as a function of $N$ and we see that the average coherence tends to an asymptotic value.

The analogous figure of merit in the continuous measurement setup of Sec.~\ref{sec:continuous} would be the average coherence w.r.t. the stationary probability distribution of the first moments. This probability distribution is the stationary solution of the Fokker-Planck equation associated with the Wiener process in Eq.~\eqref{eq:r_wiener}.

\bibliography{GCoh}
\end{document}